\documentclass[,final]{aipproc}
\layoutstyle{8x11single}

\begin{document}

\title{Unusually High Metallicity Host Of The Dark LGRB 051022}


\author{J. F. Graham}{
  address={Space Telescope Science Institute}
  ,altaddress={Johns Hopkins University}}
\author{A. S. Fruchter}{address={Space Telescope Science Institute}}
\author{L. J. Kewley}{address={University of Hawaii}}
\author{E. M. Levesque}{address={University of Hawaii}}
\author{A. J. Levan}{address={University of Warwick}}
\author{N. R. Tanvir}{address={University of Leicester}}
\author{D. E. Reichart}{address={University of North Carolina at Chapel Hill}}
\author{M. Nysewander}{address={Space Telescope Science Institute}}

\begin{abstract}
We present spectroscopy of the host of GRB 051022 with GMOS nod and shuffle on Gemini South and NIRSPEC on Keck II.  We determine a metallicity for the host of log(O/H)+12 = 8.77 using the R23 method (Kobulnicky \& Kewley 2004 scale) making this the highest metallicity long burst host yet observed.  The galaxy itself is unusually luminous for a LGRB host with a rest frame B band absolute magnitude -21.5 and has the spectrum of a rapidly star-forming galaxy.  Our work raises the question of whether other dark burst hosts will show high metallicities.
\end{abstract}

\maketitle
\vspace{-0.6 in}
\subsection{Introduction}
\vspace{-0.13 in}
GRB 051022 was initially detected by the by the Fregate, WXM, and SXC instruments on the HETE-2 satellite (\citealt{GCN4131} GCN 4131).  The burst was sufficiently bright in soft X-rays that the SXC position was determined independently of the WXM location (\citealt{GCN4137} GCN 4137).  HETE observations indicate a burst duration in excess of 2 to 4 minutes securely identifying this as a Long Gamma Ray Burst (LGRB).  Approximately 3.5 hours after the initial HETE-2 detection the burst was observed with with the X-Ray Telescope on the Swift satellite and a bright uncatalogued fading X-ray source was detected at RA: 23$^{h}$56$^{m}$04.1$^{s}$ Dec: +19$^{\circ}$36$^{m}$25.1$^{s}$ (J2000) with 4" error (\citealt{GCN4141} GCN 4141) 67 arc seconds from the HETE-2 position.

Radio, Millimeter, and X-ray sub arc-second positions locate the burst on an extended object with an approximate 5:3 elongation ratio (see Figure \ref{image}) which is thus identified as the burst's host galaxy.  Spectroscopy with the Palomar Observatory 200" Hale telescope places this object at a redshift of z = 0.8 (\citealt{GCN4156} GCN 4156).  The galaxy is  unusually bright for a LGRB host (rest frame B band absolute magnitude about -21.5) and belongs at about L* on the Schechter luminosity function.  However ground-based imaging is not sufficient to discern its morphology (see figure \ref{image}).  No optical or IR fading was observed classifying this as a dark burst.


\vspace{-0.25 in}
\subsection{Host Galaxy Spectroscopy}
\begin{figure}[ht]
\includegraphics[width=1\textwidth]{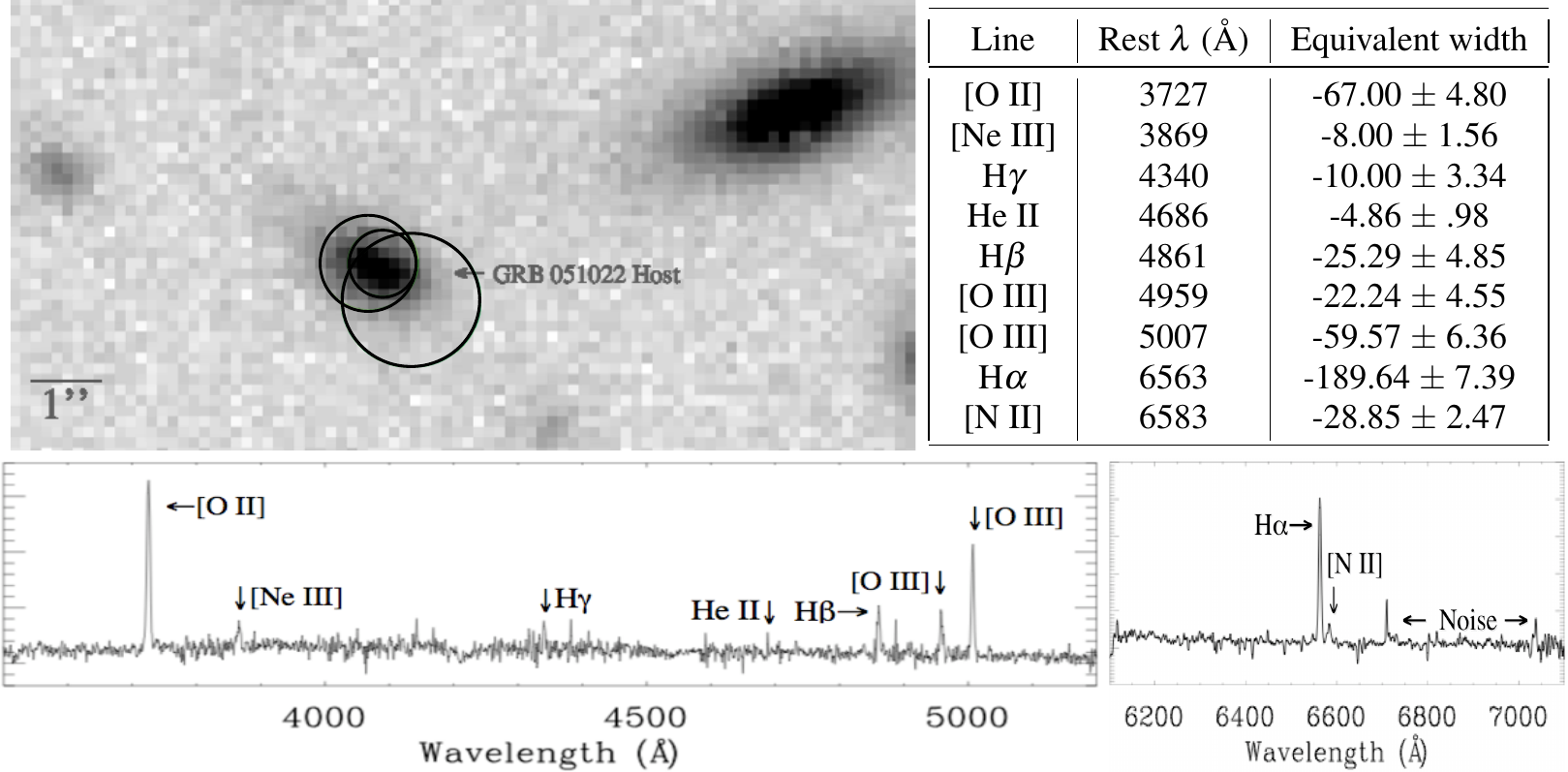}
\caption{TOP LEFT: Gemini South GMOS (r band) image of the GRB 051022 host galaxy .146 arc seconds per pixel.  GRB 051022 astrometry error circles from Chandra (0.7" \citealt{GCN4163} GCN 4163), VLA (1" \citealt{GCN4154} GCN 4154), and millimeter observations (0.5" \citealt{GCN4157} GCN 4157) overplotted.   Note that the morphology of this host is unresolved in ground-based images.
\label{image}  BOTTOM: Gemini GMOS Nod \& Shuffle (left) and Keck NIRSPEC (right) spectrum of GRB 051022 shifted to the rest frame and table of equivalent widths of spectral lines.  The [O II], [O III], and H$\beta$ lines are used for the R$_{23}$ method and the [N II] to H$\alpha$ ratio for breaking the R$_{23}$ degeneracy. \label{spectra}\label{spec}}
\end{figure}

\vspace{-0.13 in}
Initial spectroscopic observations were obtained with the GMOS instrument on Gemini North on the night of the November 25$^{th}$ 2005.  A central wavelength of 7500 {\AA} was selected yielding a spectral range of 5500 to 9500 {\AA}.  The R400 grating offers a reasonable compromise between spectral resolution (1.37 {\AA}/pixel) and width of coverage (about 4000 {\AA}) and was thus selected.  Due to the abundance of skylines in the spectral range the Nod \& Shuffle method was used.  This yielded a spectrum with a spectral resolution of 1.37 {\AA} per pixel and a spatial resolution of 0.15 arc seconds per pixel.  The host galaxy spectrum contains several bright emission lines placing it at a redshift of $z=0.806$ and allowing determination of the equivalent widths of several observed lines (see Figure \ref{spec}).

Subsequent Near infrared spectroscopy was performed with \textit{NIRSPEC} on the Keck II telescope on the night of October 23$^{th}$.  Our observations consisted of four 900 second exposures  in the NIRSPEC-3 filter, using a 0.76 arc-second slit, and giving a  spectral coverage from 1.15 to 1.33  $\mu$m and a spectral resolution  of 2.33 {\AA} per pixel.  This yielded a spectrum with a spectral resolution of 1.92 {\AA} per pixel and a spatial resolution of .16 arc seconds per pixel with a plainly visible detection of the H$\alpha$ and 6583 {\AA} [N II] lines in the resulting image and yields a value of 6.57 $\pm$ 0.52 for the H$\alpha$ to 6583 {\AA} [N II] line flux ratio.

%
%
%
%
%
%
%
%
%
%


\vspace{-0.25 in}
\subsection{Metallicity}
\vspace{-0.13 in}
The R$_{23}$ method is a commonly used metallicity diagnostic based on the electron temperature sensitivity of the oxygen spectral lines.  First proposed by Bernard Pagel in 1979, R$_{23}$ has become the primary metallicity diagnostic for galaxies at $z>0.3$ (especially those where the faint [O III] 4363 {\AA} line is not measurable).  Although it directly measures the metallicity in H II regions it is used extensively to measure the overall metallicity of other galaxies.

Using the [O II], [O III], and H$\beta$ host galaxy line equivalent widths from our Gemini South Nod \& Shuffle Spectroscopy with the R$_{23}$ method described in Kobulnicky \& Kewley 2004 we calculate two degenerate and widely differing metallicity values  -- slightly super-solar or about 1/4 solar.

To break this degeneracy we obtained direct observation of the [N II] 6583 {\AA} line and H$\alpha$ 6563 {\AA} line strengths with NIRSPEC on Keck II.  Using the ratio method from Kewley \& Ellison 2008 on those line strengths we select the upper branch of the degeneracy giving a host galaxy metallicity of log(O/H)+12 = 8.77 in the Kobulnicky \& Kewley 2004 scale (or log(O/H)+12 = 8.62 in the McGaugh 1991 scale), the highest measured metallicity of any long burst host galaxy.

\vspace{-0.25 in}
\subsection{Conclusions}
\begin{figure}[h]
\includegraphics[width=.39\textwidth,height=.39\textwidth]{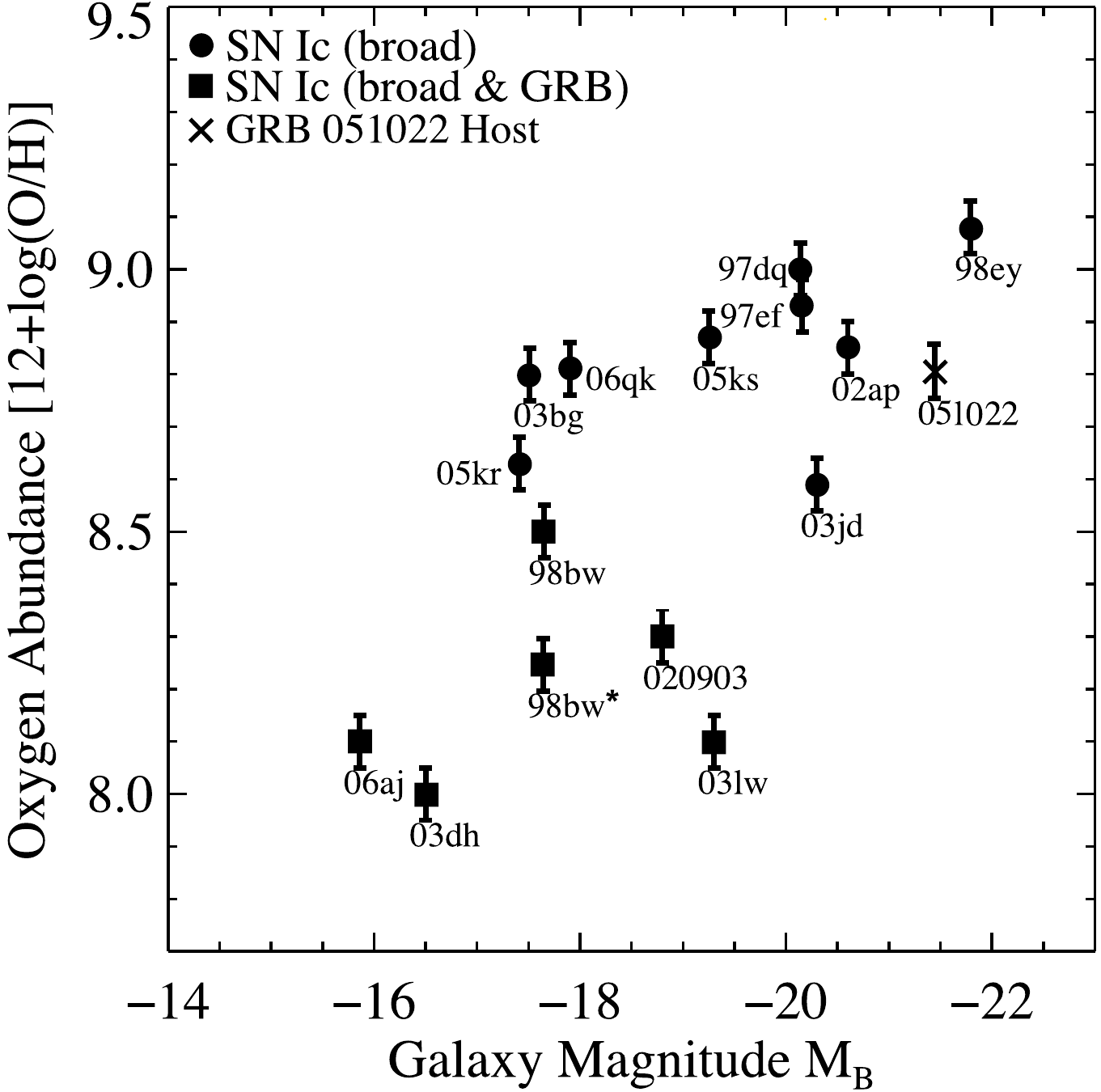}
\includegraphics[width=.39\textwidth,height=.39\textwidth]{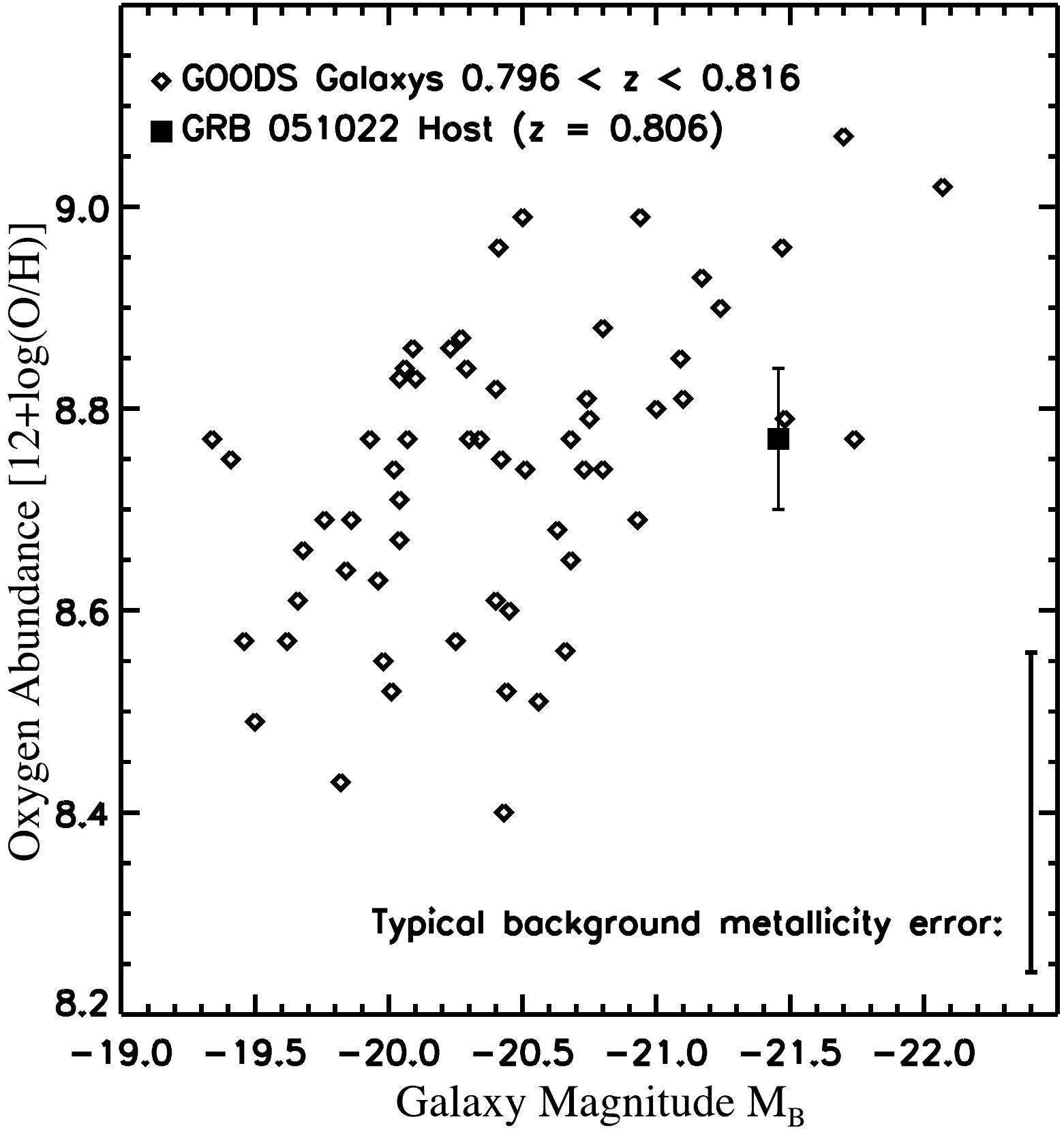}
\includegraphics[width=.2\textwidth]{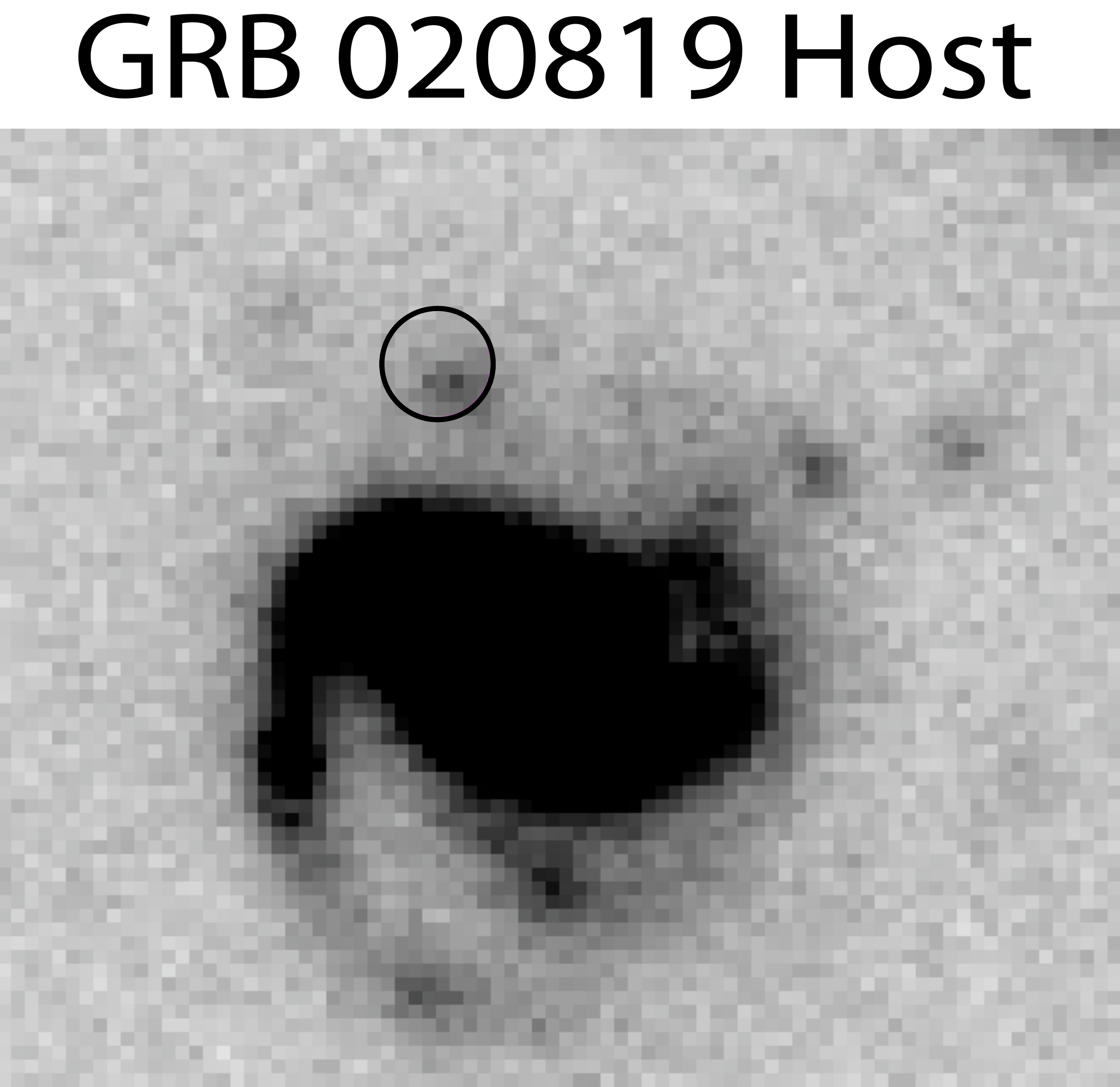}
\caption{Left \& center: LGRB 051022 has the highest measured metallicity of any long burst host. However its metallicity is similar to Ic's without GRBs (left plot adapted from figure 5 in Modjaz et al. 2008) and to comparably luminous galaxies at similar redshifts (center plot).  \label{modjaz}  RIght: Gemini North GMOS (r band) image of the GRB 020819 host galaxy.  The 1" diameter radio afterglow error circle is shown.  Being both dark bursts with similarly exceptionally bright host galaxies this object is a good analogy for GRB 051022.  If shifted out to a similar redshift and viewed from an edge on orientation it would be consistent with our Gemini ground based imaging on the GRB 051022 host. \label{020819}}
\end{figure}

\vspace{-0.13 in}
The host of GRB 051022 is among the brightest LGRB host galaxies yet seen.  Its magnitude alone makes it an unlikely host and may be related to the potential extinction of the optical counterpart and absence of a supernova event.  Our observations place the metallicity of the GRB 051022 host galaxy at log(O/H)+12 = 8.77 (in the Kobulnicky \& Kewley 2004).  While this is the highest measured metallicity of any long burst host galaxy, its metallicity is similar to IcÕs without GRBs and slightly lower but still consistent with comparably luminous galaxies at similar redshifts (see Figure \ref{modjaz}).

Such an observed high metallicity could be due to the LGRB being located in a lower metallicity region of its host then the galactic average measured.  Direct measurement of just such a case (the progenitor region denoted with an asterisk in Figure \ref{modjaz}) has been claimed by \cite{Hammer} for LGRB 080425 (SN 98bw) the previous highest metallicity LGRB host galaxy.  Unfortunately, given the limited degree of accuracy of the placement of the LGRB on itÕs host and spatial resolution in ground based spectroscopy, it is not practical to determine the metallicity of the progenitor region of the GRB 051022 host.  Also, given these constraints, the burst could be located in an outlying region of the galaxy as was the case for LGRB 020819.  Being both dark bursts with similarly exceptionally bright host galaxies of similar luminosity and linear size this object is a good analogy for GRB 051022 and were the face on orientation of the system rotated to the likely edge on orientation of the GRB 051022 host and shifted out to an equivalent redshift (from $z=0.4$ to $z=0.8$) it would be almost identical to our Gemini imaging of GRB 051022.  X-ray observations of GRB 051022 by Nakagawa et al.\thinspace (2006) estimate an optical extinction of up to 37 magnitudes (R band) making the absence of a detected optical counterpart or supernova event understandable in the case of the LGRB being located on the far edge of an inclined edge on spiral system.

Another tantalizing morphological possibility is a collision of a typical low metallicity LGRB host type galaxy with another likely higher metallicity galaxy creating an intense burst of star-formation yielding a higher luminosity to mass ratio than the galaxies posses in a static situation.  In such a situation the GRB might still be found in low metallicity gas.  A region of fainter extended emission (about 25 magnitude per square arc-second) with similar strong blue color to the rest of the object is observed (see figure \ref{020819} above and to the left of the error circles) which, combined with the photometric analysis of \cite{CastroTirado} pointing to a recent past starburst event, and our spectroscopic evidence of a high star-formation, makes such a merging system also a strong possibility.

Alternately, GRB 051022 being a dark burst, this poses interesting questions as to the nature of dark bursts and whether they share the same metallicity aversion observed in long bursts with optical afterglows, however much further study of dark burst host metallicities is necessary.

\vspace{-0.25 in}
\bibliographystyle{aipproc}
\bibliography{\jobname}
\end{document}